\documentclass[12pt]{article}
\usepackage[centertags]{amsmath}
\usepackage{amsfonts}
\usepackage{bm}
\begin{document}
\title{\text \bf{ Quantization of  Einstein Gravity:\\ 
		Extraction of Hilbert Space and Constraints}}
\author{H.S.Sharatchandra\thanks{E-mail:
{sharat@cpres.org}} \\[2mm]
{\em Centre for Promotion of Research,} \\
{\em 7, Shaktinagar Main Road, Porur, Chennai 600116, India}}
\date{}
\maketitle
\begin{abstract}
In this paper, a careful treatment of extraction of the Hilbert space and constraints from the formal functional integral with the Einstein-Hilbert action is given. The diffeomorphism inavariant measure is worked out using the metric of metrics. The procedure of quantizing the classical constraints is bypassed. Instead the relevant operator constraints are directly obtained from the functional integral, removing the ambiguities and uncertainties involved in guessing them from from the classical theory. The novelties of the resulting formalism are briefly discussed.
\end{abstract}

\smallskip

\smallskip


Dirac's prescription to quantize a classical theory is to ~i. obtain canonical variables and Hamiltonian; ii. promote canonical variables to operators and Poisson brackets to commutators. When this is applied to gauge theories, such as Maxwell's theory of electromagnetism, we encounter constraints in addition. Bergmann, Dirac and many others \cite{D} applied this to Einstein's general theory of relativity. They obtained constraints related to general coordinate invariance. These constraints appear to be  intractable. The ADM formalism \cite{adm} is more appealing for the canonical approach. But the ADM \cite{D} are also percieved to be intractable, especially in the quantized version . 
 
Since the time of Dirac, there are ways of bypassing his prescription for quantization of a classical theory. There is an a priori candidate for the quantum theory: functional integral with the action of the classical theory. This approach with the Einstein-Hilbert action leads to a 'non-renormalizable' theory, lending to further doubts regarding the straightforward use of Einstein's theory. The issue has been addressed by many authors over  decades. It has led to many new proposals and theories.

In this paper, I give a careful treatment of extracting the Hilbert space and operator constraints from the formal functional integral. This resolves the first set  of the expected problems. I will show elsewhere \cite{qcI,qcII} that the issues related to quantized constraints can be handled.

Here we consider only pure gravity. We can also incorporate matter fields easily.

We are interested in tackling the partition function
\begin{equation}\label{z}
Z=\int D g\ e^{iS[g]/{\hbar}},\\
\end{equation}
with the Einstein-Hilbert action:
\begin{equation}\label{eh}
S = \frac{1}{2\kappa}\int d^{4}x \sqrt{-g(x)}R(x),
\end{equation}
where $\kappa=8\pi G c^{-4}$, $G$ is the Newton's constant.
We have to integrate over all valid metrics $g_{\alpha \beta}(x)$ over space-time labelled by coordinates $x=\{x_{\alpha}\}$.
Here $\alpha, \beta= 0,1,2 ~ or ~ 3$ and we choose the signature $(-+++)$ for the metric.

We first compute integration measure $Dg$ consistent with diffeomorphism invariance using the metric of metrics. For a infinitesimal metric $\delta g_{\alpha \beta}(x)$ the diffeomorphism invariant metric is
\begin{eqnarray}\label{mm}
<\delta g, \delta g>=\int d^4 x\sqrt{-g}\, \delta g_{\alpha \beta}g^{\alpha \gamma} g^{\beta \delta}\, \delta g_{\gamma \delta}(x).
\end{eqnarray}
This gives the formal diffeomorphism invariant functional measure
\begin{eqnarray}\label{m}
D g= \prod_{{\alpha \beta},x}dg_{\alpha \beta}(x)|det M(x)|^{1/2},
\end{eqnarray}
which is an integration over the 10 components of the metric $g_{\alpha\beta}(x)$ consistent with the signature and $M$ is the $10\times 10$ matrix
\begin{eqnarray}\label{M}
M^{\alpha\beta,\gamma \delta}=\frac{1}{2}\sqrt{-g}(g^{\alpha\gamma} g^{\beta \delta}+g^{\alpha\delta}g^{\beta\gamma}).
\end{eqnarray}
Consider the (non-covariant) eigenvalue equation for the $4\times 4$ real symmetric matrix $g^{\alpha\gamma}$,
\begin{eqnarray}\label{evg}
g^{\alpha\beta }\xi_A^{\beta}=\lambda_A \xi_A^{\alpha},
\end{eqnarray}
with real eigenvalues $\lambda_A, A=0,1,2,3$. Then 
\begin{eqnarray}\label{evM}
M^{\alpha\beta,\gamma \delta}(\xi_A^{\gamma}\xi_B^{\delta}+\xi_A^{\delta}\xi_B^{\gamma})=\sqrt{-g}\lambda_A \lambda_B (\xi_A^{\alpha}\xi_B^{\beta}+\xi_A^{\beta}\xi_B^{\alpha}).
\end{eqnarray}
Thus $M$  has  $10$ eigenvalues  $\sqrt{-g}\lambda_A\lambda_B, A,B=0,1,2,3, ~ A \leq B$. Now 
$g=det(g_{\alpha\beta})=1/det(g^{\alpha\beta})
=\prod_A 1/\lambda_A$. Therefore    
\begin{eqnarray}\label{detM}
|det M|=(\sqrt{-g})^{10}\prod_{A,B; A \leq B} (\lambda_A\lambda_B)
=(\sqrt{-g})^{10}(\prod_{A} \lambda_A)^5=1.
\end{eqnarray}
Thus the diffeomorphism invariant measure is simply
\begin{eqnarray}\label{m1}
D g= \prod_{{\alpha \beta},x}dg_{\alpha \beta}(x).
\end{eqnarray}
If we had repeated these steps for the contravariant $g^{\alpha\beta}$ we would have ended up with 
\begin{eqnarray}\label{m2}
D g= \prod_{{\alpha \beta},x}dg^{\alpha \beta}(x) (\sqrt{-g(x)})^{10}.
\end{eqnarray}
The simplicity of the measure Eqn.\ref{m1} is unique to 4-dimensional space-time. In 3-dimensional space-time we get 
\begin{eqnarray}\label{m3}
d=3: D g= \prod_{{\alpha \beta},x}\frac{dg_{\alpha \beta}(x)} {\sqrt{-g(x)}}.
\end{eqnarray}

Now we use ADM variables \cite{adm}, \cite{D}  for the metric:
\begin{equation}\label{adm}
{g_{\alpha\beta} = \left( \begin{array}{cc}
	g_{00} & g_{0b} \\
	g_{a0} & g_{ab}
	\end{array}  \right) =
	\left( \begin{array}{cc}
	-N^2+{\cal N}_a q^{ab} {\cal N}_b & {\cal N}_b \\
	{\cal N}_a & q_{ab}
	\end{array} \right) }.
\end{equation}
Here $q_{ab}, a,b=1,2~ or ~3$ are the renamed spatial components of the 
metric $g_{\alpha\beta}$ and $q^{ab}$ is its inverse. $N$, ${\cal N}_a$ are respectively called the lapse and shift functions. Components of the contravariant metric $g^{\alpha\beta}$ are given by the matrix inverse,
\begin{equation} \label{adm1}
{ g^{\alpha\beta} = \left( \begin{array}{cc}
	g^{00} & g^{0b} \\
	g^{a0} & g^{ab}
	\end{array} \right) =
	\left( \begin{array}{cc}
	-1/N^2 & {\cal N}^b/N^2 \\
	{\cal N}^a/N^2 & q^{ab}-{\cal N}^a {\cal N}^b/N^2
	\end{array} \right) }.
\end{equation}
${\cal N}^a$ is obtained from ${\cal N}_a$ by raising the index using $q^{ab}$. We get 
\begin{equation} \label{detg1}
g=-N^2 q,
\end{equation}
where $q=det(q_{ab})$. With this change of variables we have
\begin{eqnarray}\label{m5}
\prod_{{\alpha \beta}}dg_{\alpha \beta}=
2NdN\prod_{ab}dq_{ab}\prod_{a}d{\cal N}_a.
\end{eqnarray}
The Einstein-Hilbert action in terms of the ADM variables is
\begin{eqnarray}\label{ehadm}
S[g]=\frac{1}{2\kappa}\int d^4 x N\sqrt{q}(K_{ab}G^{abcd}K_{cd}+R^{(3)})(x).
\end{eqnarray}
Here $R^{(3)}$ is the intrinsic curvature of the hypersurface $x_0=constant$ and $K_{ab}$ is its extrinsic curvature,
\begin{eqnarray}\label{K}
K_{ab}=\frac{1}{2N}(\dot{q}_{ab}-D_a{\cal N}_b-D_b{\cal N}_a).
\end{eqnarray}
Also
\begin{eqnarray}\label{G}
G^{abcd}=\frac{1}{2}(q^{ac}q^{bd}+q^{ad}q^{bc}-2 q^{ab}q^{cd}),
\end{eqnarray}
is a $6 \times 6$ matrix called the DeWitt tensor (up to a $\sqrt{q}$ factor). 
(In this paper we ignore all boundary terms by presuming relevant boundary conditions. It is possible to consider the effects of boundary terms also in our analysis.)

We use these variables in Eqn.\ref{z}.
We linearize terms quadratic in $K_{ab}$ by using the master formula
\begin{eqnarray}\label{mf}
exp(iq_{a}G^{ab}q_{b})=|det(G^{-1})^{ab}|^{1/2}\int \Pi_{a} dp^{a} exp(i(2q_{a}p^{a}-p^{a}(G^{-1})^{ab}p^{b})).
\end{eqnarray}
For us,
\begin{eqnarray}\label{mf1}
exp(i\frac{N\sqrt{q}}{2\kappa \hbar}K_{ab}G^{abcd}K_{cd})=|det(\frac{2\kappa \hbar }{N\sqrt{q}}G_{abcd})|^{1/2}\int \Pi_{ab} dp^{ab}\\ \nonumber exp(i(\frac{2N}{\hbar}K_{ab}p^{ab}-p^{ab}\frac{2\kappa N}{\hbar\sqrt{q}}G_{abcd}p^{cd}),
\end{eqnarray}
where 
\begin{eqnarray}\label{G1}
G_{abcd}=\frac{1}{2}(q_{ac}q_{bd}+q_{ad}q_{bc}-q_{ab}q_{cd}),
\end{eqnarray}
is the inverse of the DeWitt tensor,
\begin{eqnarray}\label{G2}
G_{abcd}G^{cdef}=\frac{1}{2}(\delta_{ae}\delta_{bf}+\delta_{af}\delta_{be}).
\end{eqnarray}
Note that the eigenfunctions of $G$ are not as simple (Eqn.\ref{evM}) as those of $M$ in terms of the eigenfunctions of $q$. Nevertheless the determinant \cite{D} is simply a power of $q$: $det (G_{abcd})\sim |q|^{4}$.Therefore
\begin{eqnarray}\label{detG2}
det(\frac{2\kappa \hbar}{N\sqrt{q}}G_{abcd})\sim  (\frac{2\kappa \hbar}{N\sqrt{q}})^6 q^4 \sim \frac{q}{N^6}.
\end{eqnarray}
Using all these we get
 \begin{eqnarray}\label{ehadm2}
 Z(C) &\sim &\int \prod_{ab,x}dp^{ab}(x)\prod_{ab,x}dq_{ab}(x) \prod_{x}
 dN(x)\prod_{a,x}
d{\cal N}_a(x) \, \frac{\sqrt{q}}{N^2}\\ \nonumber
 && exp(\frac{i}{\hbar}\int d^4x (2NK_{ab}p^{ab}-N(\frac{2\kappa}{\sqrt{q}}p^{ab}G_{abcd}p^{cd}
  -\frac{\sqrt{q}}{2\kappa}R^{(3)}))(x)).
 \end{eqnarray}
This has the canonical form for a Hamiltonian interpretation
\begin{eqnarray}\label{ehadm3}
Z(C)&=&\int \prod_{ab,x}dp^{ab}(x)\prod_{ab,x}dq_{ab}(x) \prod_{x}
 dN(x)\prod_{a,x}d{\cal N}_a(x)\, \frac{\sqrt{q}}{N^2}\\ \nonumber
&&exp(\frac{i}{\hbar}\int d^4x(p^{ab}{\dot q}_{ab} -N(\frac{2\kappa}{\sqrt{q}}p^{ab}G_{abcd}p^{cd}
-\frac{\sqrt{q}}{2\kappa}R^{(3)})+2{\cal N}_aD_{b}p^{ab})(x)).
\end{eqnarray}
Note the following:

i. If we use Feynman's time slicing procedure, we get canonically conjugate fields with equal time commutation rules
\begin{eqnarray}\label{ccr}
[p^{ab}(X,t),q_{cd}(Y,t)]=-\frac{i\hbar}{2}\delta^3(X-Y)(\delta_{ac} \delta_{bd} +\delta_{ad} \delta_{bc}),
\end{eqnarray}
with other commutators being zero. Here $X,Y$ etc. stand for spatial coordinates. We have used definitions to avoid invariant densities $\sqrt{q(X)}\delta^3(X-Y)$ which can cause operator ordering problems later.

ii. Canonical conjugates of fields $N(x),{\cal N}_a(x)$ do not appear in Eqn.~\ref{ehadm3}. They are playing the role of Lagrange multipliers.

iii. At this level the relevant Hilbert space basis is formally
$|\{q_{ab}(X),N(X),{\cal N}_a(X)\}>$ or equivalently $|\{p^{ab}(X),N(X),{\cal N}_a(X)\}>$ with the inner product
\begin{eqnarray}\label{ip}
<\{q_{ab}(X),N(X),{\cal N}_e(X)\}|\{p^{cd}(Y),N'(Y),{\cal N'}_f(Y)\}> \\ \nonumber =exp(\frac{i}{\hbar}\int d^3X p^{ab}(X)q_{ab}(X))
\prod_X \delta (N(X)-N'(X))\prod_{a,X} \delta ({\cal N}_a(X)-{\cal {N'}}_a(X)).
\end{eqnarray}

We can handle the 'cyclic coordinates' or 'ignorable fields' $N, {\cal N}_a$
in different ways as in gauge theories.

Version I: 'Fix the gauge', $N(x)=1,{\cal N}_a(x)=0$. This corresponds to $g_{00}=-1, g_{0a}(x)=0$. This is consistent with the signature of the metric. We get a conventional type of functional integral,
\begin{eqnarray}\label{ehadm5}
Z(C)&=&\int \prod_{ab,x}dp^{ab}(x)\prod_{ab,x}dq_{ab}(x) \sqrt {q(x)}\\ \nonumber
&&exp(\frac{i}{\hbar}\int d^4x(p^{ab}(x){\dot q}_{ab}(x)
 -(\frac{2\kappa}{\sqrt{q(x)}}p^{ab}(x)G_{abcd}(x)p^{cd}(x)
-\frac{\sqrt{q(x)}}{2\kappa}R^{(3)}(x))).
\end{eqnarray}

Version II: We are more interested in getting a formulation close to the classical case where we get the 'momentum' and 'Hamiltonian' constraints. Integrating over ${\cal N}_a(x)$ we get a functional delta function,
\begin{eqnarray}\label{mc}
\prod_{x}\delta (D_{b}p^{ab}(x)),
\end{eqnarray}
which corresponds to the momentum constraint of the ADM formalism. 

The field $N(x)$ has to be treated differently. It cannot be formally integrated over the range $(-\infty,+\infty)$ as this is not consistent with the signature of the metric. In addition the measure is formally $N^{-2}dN$ and not just $dN$. However we have the freedom to choose the gauge $N \rightarrow \infty$, which is consistent with the signature of the metric for any values of $q_{ab},{\cal N}_a$. Then we again get a delta functional in the functional integral:
\begin{eqnarray}\label{hc}
\prod_{x}\delta (\frac{2\kappa}{\sqrt{q(x)}}p^{ab}(x)G_{abcd}(x)p^{cd}(x)
-\frac{\sqrt{q(x)}}{2\kappa}R^{(3)}(x)),
\end{eqnarray}
which corresponds to the Hamiltonian constraint of the  ADM formalism.

Thus the functional integral with the Einstein-Hilbert action is formally equivalent to
\begin{eqnarray}\label{ehadm5}
Z(C)&=&\int \prod_{ab,x}dp^{ab}(x)\prod_{ab,x}dq_{ab}(x) \sqrt{q(x)}\,
exp(\frac{i}{\hbar}\int d^4x p^{ab}(x){\dot q}_{ab}(x))\\ \nonumber
&&\prod_{a,x}\delta (D_{b}p^{ab}(x))
\prod_{x}\delta (\frac{2\kappa}{\sqrt{q(X)}}p^{ab}(x)G_{abcd}(x)p^{cd}(x)
-\frac{\sqrt{q(X)}}{2\kappa}R^{(3)}(x)).
\end{eqnarray}
We want to obtain the Hilbert space interpretation of the Dirac delta functionals. Feynman's time slicing gives the formal discretization
\begin{eqnarray}
exp(\frac{i}{\hbar}\int d^4x p^{ab}(x){\dot q}_{ab}(x)) \sim
\int \prod_{ab,X,m}dp^{ab}(X,t_m)dq_{ab}(X,t_m) \\ \nonumber
\cdots<\{q_{ab}(Y,t_{n+1})\}|\{p^{ab}(Y,t_{n})\}>
<\{p^{ab}(Y,t_{n})\}|\{q_{ab}(Y,t_{n})\}>\cdots.
\end{eqnarray}
Consider the Dirac delta functionals at one time,
\begin{eqnarray}
<\{q_{ef}(Z)\}|\prod_{X}\delta (D_{b}p^{ab}(X))|\{p^{ef}(Z)\}>
=<\{q_{ef}(Z)\}|\int dN_a(X)e^{i\int d^3X N_a(X) D_{b}p^{ab}(X)/\hbar}|\{p^{ef}(Z)\}>.
\end{eqnarray}
To get an operator interpretation we generalize Feynman's time slicing procedure. Any given $N_a(X)$ is sliced into a large number $M$ of infinitesimal bits of $N_a(X)/M$. For each we have the approximation
\begin{eqnarray}
<\{q_{ef}(Z)\}|1+\frac{i}{\hbar}\int d^3X \frac{N_a(X)}{M} D_{b}p^{ab}(X)|\{p^{ef}(Z)\}>.
\end{eqnarray}
Keeping $N_c(X)$ as a c-number field at present, we can interprete this as the matrix element $<\{q\}|1+\hat O/M|\{p\}>$ where the operator
\begin{eqnarray}
\hat O=\int d^3Y  N_b(Y) \hat D_{a}\hat p^{ab}(Y). \label{o}
\end{eqnarray}
Here $\hat D_{a}$ is the covariant derivative with the metric $ q_{ab}$ is replaced by the operator field  $\hat q_{ab}$ and all such fields 
in $\hat D_{a}$ are on the left of $\hat p^{ab}$ in Eqn.\ref{o}. 	Now
\begin{eqnarray}
[\frac{i}{\hbar}\hat O, \hat q_{ab}(X)]=-\hat D_{a}N_b(X)-\hat D_{b}N_a(X),
\end{eqnarray}
which can be recognized as the transformation of  the metric under an  infinitesimal diffeomorphism $\delta X^a=N^a(X)$,
\begin{eqnarray}
\delta q_{ab}(X)=-D_{a}N_b(X)-D_{b}N_a(X).
\end{eqnarray}
This looks like a highly non-linear and inhomogenious transformation, but it is not so because
\begin{eqnarray}
D_{a}N_b(X)+D_{b}N_a(X)=\partial_{a}N^c(X) q_{cb}+\partial_{b}N^c(X) q_{ca}-N^c(X)\partial_{c}q_{ab}. \label{l}
\end{eqnarray}
This is the infinitesimal version  (with $X'^a=X^a+N^a(X)$) of the general coordinate transformation
\begin{eqnarray}
q'_{ab}(X')=\frac{\partial X^c}{\partial X'^a}\frac{\partial X^d}{\partial X'^b}q_{cd}(X),
\end{eqnarray}
which is a linear and homogeneous transformation.
Using Eqn.\ref{l} in $\hat O$ and presuming $N^a$ (and therefore  $N_a=q_{ab}N^b$ do not) commute with $p^{ab}$, we also get the transformation of $\hat p^{ab}$,
\begin{eqnarray}
[\frac{i}{\hbar}\hat O, \hat p^{ab}(X)]=\partial_{c}N^a(X) \hat p^{cb}(X)+\partial_{c}N^b(X) \hat p^{cb}(X)-\partial_{c}(N^c(X) \hat p^{ab}(X)).\label{p}
\end{eqnarray}
Note that with our equal time commutation rules \ref{ccr}
we require $p^{ab}$ to transform as a symmetric tensor of weight one:
\begin{eqnarray}
p'^{ab}(X')=det(\frac{\partial X'^e}{\partial X^f})^{-1}\frac{\partial X'^a}{\partial X^c}\frac{\partial X'^b}{\partial X^d}p^{cd}(X).
\end{eqnarray}
Eqn.\ref{p} is exactly the infinitesimal version of this.

Thus we have demonstrated that $\hat O$ (with the stated ordering of the operators and $\hat N^a$ (but not $\hat N_a$) commuting with $\hat{q}_{ab}(X),\hat{p}^{ab}(X)$) is the generator of infinitesimal diffeomorphism transformation of field operators $\hat{q}_{ab}(X),\hat{p}^{ab}(X)$. Thus
\begin{eqnarray}
<\{q_{ef}(Z)\}|(exp(\frac{i}{M\hbar}\int d^3X N_a(X) D_{b}p^{ab}(X))^M|\{p^{ef}(Z)\}>\\ \nonumber
=<\{q_{ef}(Z)\}|(exp(\frac{i}{M\hbar}\int d^3X N_a(X)D_{b}p^{ab}(X)))^{M-1}|\{p^{ef}(Z)\}_{N/M}>,
\end{eqnarray}
where $\{p^{ef}(Z)\}_{N/M}$ means infinitesimal diffeomorphism of $\{p^{ef}(Z)\}$ by $N^a(X)/M$. Recurrence of this procedure $M$ times in infinitesimal steps each of $N^a(X)/M$ gives
\begin{eqnarray}
e^{i\int d^3X N^a(X) \hat D_{b}\hat p_a^b(X)}|\{p^{ef}(Z)\}>
=|\{p^{ef}(Z)\}_{N}>,
\end{eqnarray}
which is a finite diffeomorphism transformation corresponding to  $X'^a=X^a+N^a(X)$.
The operator $\hat D_{b}\hat p_a^b(X)$ is not formally self-adjoint. We rectify this by replacing it by 
\begin{eqnarray}
\hat P_a(X)=\frac{1}{2}(\hat D_{b}\hat p_a^b(X)+\hat p_a^b(X)\hat D_{b}^L),
\end{eqnarray}
where $\hat D_{b}^L$ is again the covariant derivative with the metric replaced with the operator field $\hat q_{ef}$, and in addition the ordinary derivative $\nabla$ acting on the left (on $\hat p^{ab}$). This does not alter the equations above. $\hat P_a(X), a=1,2,3$ are the generators of  infinitesimal 3-diffeomorphisms. The operator 
\begin{eqnarray}
P=\int DN^a(X) e^{i\int d^3X N^a(X) \hat P_a(X)},
\end{eqnarray}
is a projection operator serving to average over all diffeomorphism transformations on any state. This way the 'physical states' are invariant under diffeomorphism transformations. (This is analogous to gauge theories where an integration over $A_0$ in the functional integral gives the Gauss law constraint.) This also means that  we have to use only observables which are invariant under diffeomorphism transformations, which commute with the $ P$, and the completeness relations are now modified to 
\begin{eqnarray}
1=\int Dq_{ab}(X)
P|\{q_{ab}(X)\}><\{q_{ab}(X)\}|P, \label{di}
\end{eqnarray}
with a similar identity involving $\{p^{ab}(X)\}$ also.

Now we address the meaning of 
\begin{eqnarray}
\prod_{X}<\{p^{ab}(Y)\}|\delta (\frac{2\kappa}{\sqrt{q(X)}}p^{ab}(X)G_{abcd}(X)p^{cd}(X)
-\frac{q(X)}{2\kappa}R(X))|\{q_{ab}(Y)\}>.
\end{eqnarray}
We use procedure followed above. To get a self adjoint operator \cite{D}, we interprete this as 
\begin{eqnarray}
\int dN(X,t_{n})(exp(i\int d^3X \frac{N(X,t_{n})}{M}(p^{ab}\frac{2\kappa}{\sqrt{q}}
G_{abcd}p^{cd}
-\frac{\sqrt{q}}{2\kappa}R^{(3)})(X,t_{n})))^{M-1}\\ \nonumber <\{q_{ab}(Y,t_{n+1})\}|(1+i\int d^3X \frac{N(X)}{2M}(\frac{2\kappa}{\sqrt{\hat q}}\hat G_{abcd}\hat p^{ab}\hat p^{cd}
-\frac{\sqrt{\hat q}}{2\kappa}\hat R^{(3)})(X)|\{p^{ab}(Y,t_{n})\}>\\ \nonumber
<\{p^{ab}(Y,t_{n})\}|(1+i\int d^3X \frac{N(X)}{2M}(\hat p^{ab}\hat p^{cd}\hat G_{abcd}\frac{2\kappa}{\sqrt{\hat q}}
-\frac{\sqrt{\hat q}}{2\kappa}\hat R^{(3)})(X))|\{q_{ab}(Y,t_{n})\}>.
\end{eqnarray}
Using the completeness relation for $\{p^{ab}(Y,t_{n})\}$, we get the operator
\begin{eqnarray}
\large P'=\int DN(X) e^{i\int d^3X N(X) \hat  H(X)},
\end{eqnarray}
sandwiched beteween $<\{q_{ab}(Y,t_{n+1})\}|$ and $|\{q_{ab}(Y,t_{n})\}>$.
Here
\begin{eqnarray}
\hat  H(X)=\frac{1}{2}(\frac{2\kappa}{\sqrt{\hat q(X)}}\hat G_{abcd}(X)\hat p^{ab}(X)\hat p^{cd}(X)+\hat p^{ab}(X)\hat p^{cd}(X)\hat G_{abcd}(X)\frac{2\kappa}{\sqrt{\hat q(X)}})
-\frac{\sqrt{\hat q(X)}}{2\kappa}\hat R^{(3)}(X).
\end{eqnarray}
We consider the algebra of operators \cite{D} $\hat P_a(X),\hat H(X)$ in \cite{qcII}. On diffeomorphism invariant states Eqn.\ref{di}, $\hat H(X)$ commute at different space points. Therefore they can be simultaneously diagonalized and we can write
\begin{eqnarray}
P'=\Pi_X\delta (\hat  H(X)).
\end{eqnarray}
Only states annihilated (i.e. of eigenvalue zero) by the operator $\hat H(X)$ for each $X$ are physical states, in the sense that only such states  contribute. This is the Hamiltonian constraint of quantum gravity: only states of zero energy density contribute. In other field theories there is only the ground state in classical theory with such a property. As the Hamiltonian is not positive definite \cite{D}, in case of Einstein gravity such states are numerous. Therefore the physical properties in Einstein gravity are governed by the distribution of the density of states appropriate for those properties.

There are serious problems in defining products of operators at the same space-time point in quantum field theory. In addition we have products of non-commuting operators $\hat q_{ab}(X,t),\hat p^{ab}(X,t)$ at the same space time point. Therefore the operators $\hat H(X)$ appear to be intractable. We develop techniques to handle this in \cite{qcI,qcII}.

Note that factors of $i=\sqrt{-1}$ has played a crucial role throughout. Removing them by 'Euclideanization' causes havoc in getting a meaningful interpretation. In this sense gravity is closer to Chern-Simons theories. We will discuss this issue in a greater detail elsewhere.

In spite of the crucial role played by $i=\sqrt{-1}$ throughout, the result is closer to the microcanonical ensemble of a statistical problem. This seems to be  the reason behind a close relationship between blackhole physics and thermodynamics.

It may be argued using  Version I above that there is a Hamiltonian and states of non-zero energy also contribute to the functional integral. But this can be viewed as a gauge artifact as the lapse field $N$ is not invariant under diffeomorphisms involving time also. We use Version II as it is closer to classical case and has more appealing interpretations. This also begs the question: Where are dynamics and time correlations? We address this question elsewhere.


\end{document}